\documentclass[twocolumn]{aastex61}
\usepackage{booktabs}
\usepackage{amsmath}
\usepackage[applemac]{inputenc}

\newcommand\aastex{AAS\TeX}

\newcommand{\Lcop}{$L_{\mathrm{CO(1-0)}}^{\prime}$}

\newcommand{\LIR}{$L_{\rm IR}$}
\newcommand{\aco}{$\alpha_{\rm CO}$}

\received{May 17, 2018}
\accepted{Oct 23, 2018}
\shorttitle{\aastex\ + molecular gas in CLJ1001 at $z$=2.51}
\shortauthors{Wang et al.}

\begin{document}

\title{Revealing environmental dependence of molecular gas content in a distant X-ray cluster at $\MakeLowercase{z}=2.51$}

\correspondingauthor{Tao Wang}
\email{taowang@ioa.s.u-tokyo.ac.jp, twang.nju@gmail.com}

\author[0000-0002-2504-2421]{Tao Wang}
\affil{Institute of Astronomy, Graduate School of Science, The University of Tokyo, 2-21-1 Osawa, Mitaka, Tokyo 181-0015, Japan}
\affil{Laboratoire AIM-Paris-Saclay, CEA/DSM/Irfu,
CNRS, Universit\'e Paris Diderot, Saclay, 
pt courrier 131, 91191 Gif-sur-Yvette, France}
\affil{National Astronomical Observatory of Japan, Mitaka, Tokyo
181-8588, Japan}

\author{David Elbaz}
\affil{Laboratoire AIM-Paris-Saclay, CEA/DSM/Irfu,
CNRS, Universit\'e Paris Diderot, Saclay, 
pt courrier 131, 91191 Gif-sur-Yvette, France}

\author{Emanuele Daddi}
\affil{Laboratoire AIM-Paris-Saclay, CEA/DSM/Irfu,
CNRS, Universit\'e Paris Diderot, Saclay, 
pt courrier 131, 91191 Gif-sur-Yvette, France}

\author[0000-0001-9773-7479]{Daizhong Liu}
\affil{Max Planck Institute for Astronomy, K{\"o}nigstuhl 17, D-69117 Heidelberg, Germany}

\author{Tadayuki Kodama}
\affil{Astronomical Institute, Tohoku University, Aramaki, Aoba-ku, Sendai 980-8578, Japan}

\author{Ichi Tanaka}
\affil{Subaru Telescope, National Astronomical Observatory of Japan, National Institutes of Natural Sciences, 650 North A'ohoku Place, Hilo, HI 96720, U.S.A.}

\author{Corentin Schreiber}
\affil{Leiden Observatory, Leiden University, NL-2300 RA Leiden, The Netherlands}

\author{Anita Zanella}
\affil{European Southern Observatory, Karl Schwarzschild Str. 2, D-85748 Garching, Germany}

\author{Francesco Valentino}
\affil{Dark Cosmology Center, Niels Bohr Institute, University of Copenhagen, Juliane Maries Vej 30, DK-2100 Copenhagen, Denmark}

\author[0000-0003-1033-9684]{Mark Sargent}
\affil{Astronomy Centre, Department of Physics and Astronomy, University of Sussex, Brighton BN1 9QH, UK}

\author{Kotaro Kohno}
\affil{Institute of Astronomy, Graduate School of Science, The University of Tokyo, 2-21-1 Osawa, Mitaka, Tokyo 181-0015, Japan}

\author{Mengyuan Xiao}
\affil{Key Laboratory of Modern Astronomy and Astrophysics in Ministry of Education, School of Astronomy and Space Sciences,
Nanjing University, Nanjing, 210093, China}

\author[0000-0003-3738-3976]{Maurilio Pannella}
\affil{Faculty of Physics, Ludwig-Maximilians-Universit{\"a}t, Scheinerstr. 1, 81679 M{\"u}nchen, Germany}

\author{Laure Ciesla}
\affil{Laboratoire AIM-Paris-Saclay, CEA/DSM/Irfu,
CNRS, Universit\'e Paris Diderot, Saclay, 
pt courrier 131, 91191 Gif-sur-Yvette, France}

\author{Raphael Gobat}
\affil{Instituto de F\'isica, Pontificia Universidad Cat\'olica de Valpara\'iso, Casilla 4059, Valpara\'iso, Chile}

\author{Yusei Koyama}
\affil{Subaru Telescope, National Astronomical Observatory of Japan, National Institutes of Natural Sciences, 650 North A'ohoku Place, Hilo, HI 96720, U.S.A.}

\begin{abstract}

We present a census of the molecular gas properties of galaxies in the most distant known X-ray cluster, CLJ1001, at z=2.51, using deep observations of CO(1-0) with JVLA. 
In total 14 cluster members with $M_{*} > 10^{10.5} M_{\odot}$ are detected, including all the massive star-forming members within the virial radius, providing the largest galaxy sample in a single cluster at $z > 2$ with CO(1-0) measurements. 
We find a large variety in the gas content of these cluster galaxies, which is correlated with their relative positions (or accretion states), with those closer to the cluster core being increasingly gas-poor. Moreover, despite their low gas content, the galaxies in the cluster center exhibit an elevated star formation efficiency (SFE=SFR/$M_{\rm gas}$) compared to field galaxies, suggesting that the suppression on the SFR is likely delayed compared to that on the gas content. Their gas depletion time is around $t_{\rm dep} \sim 400$ Myrs, comparable to the cluster dynamical time. This implies that they will likely consume all their gas within a single orbit around the cluster center, and form a passive cluster core by $z\sim2$.  
This result is one of the first direct pieces of evidence for the influence of environment on the gas reservoirs and SFE of $z > 2$ cluster galaxies, thereby providing new insights into the rapid formation and quenching of the most massive galaxies in the early universe. 

\end{abstract}
%\clearpage
\section{Introduction}

Galaxy clusters in the present-day universe are dominated by a population of massive, and quiescent galaxies in their center~\citep{Dressler:1997}. 
The formation mechanisms of these massive galaxies and the influence environment plays in this process remain open questions. These issues are difficult to address in the local universe, as most of the massive galaxies have already been in place for 10 Gyrs and signatures of their formation history have been largely erased. Contrary to mature clusters at low redshifts, a significant population of (proto)clusters with active star formation has been found at $z > 2-4$, the peak formation epoch of massive cluster galaxies~\citep{Thomas:2005}. With a large number of massive star-forming galaxies (SFGs) in a cluster-like environment, these structures provide ideal laboratories to explore the environmental dependence of massive galaxy formation.

The different properties, e.g., star formation rates (SFR), of galaxies in $z > 2$ (proto-)clusters and field have been extensively studied. 
While a higher fraction of quiescent galaxies in dense environments is well established, no significant difference has been found on the average SFR of star-forming galaxies, i.e., the normalization of the star-forming main sequence (MS), at $z \sim 2$~\citep{Koyama:2013,Shimakawa:2017}.  A few studies show an enhanced fraction of starburst galaxies in  (proto-)clusters at both the bright~\citep{WangT:2016b,Casey:2016} and faint end of the stellar mass function~\citep{Hayashi:2016}, however, statistical samples are still required to confirm these findings. Overall, this lack of strong environmental dependence of star formation may indicate that there is significant delay between the first infall of cluster galaxies and substantial reduction in their star formation rates (SFR), which likely only take place close to the cluster core~\citep{Wetzel:2013}.

While it is not yet fully clear how SFR of star-forming galaxies depends on environment, ample evidence exists for the deficit of cold gas, the fuel of star formation, for galaxies in dense environments~\citep{Boselli:2006}. Both neutral hydrogen (HI), molecular gas, and even dust, in cluster galaxies can be severely impacted by their local environment through, e.g., ram-pressure stripping and tidal stripping, at least in the vicinity of the cluster core~\citep{Cortese:2010,Davis:2013,Jachym:2014,Jachym:2017}. On the other hand, while pioneering studies on molecular gas content in high-z clusters ($z \sim 1.5-2$) have been recently performed, there is still no consensus on the influence of environment on galaxies' gas content~\citep{Aravena:2012,Wagg:2012,Casasola:2013,Stach:2017,Noble:2017,Coogan:2018}. This is mainly driven by the limited number of detections, which is often biased towards the most gas-rich members except for a few cases~\citep{Rudnick:2017,Hayashi:2017}. In addition, most of these targeted clusters appear to be already dominated by massive quiescent galaxies in the core, but little is known about their gas and star formation properties during the epoch of their formation/quenching. Though studies of molecular gas properties in dense environments exist at $z > 2$, they only detected some of the brightest member galaxies~\citep[e.g., ][and references therein]{Tadaki:2014,LeeM:2017,Dannerbauer:2017}, inhibiting a comprehensive understanding of the gas content of cluster galaxies.

In this paper, we present a census of molecular gas properties of 14 massive SFGs in the most distant known X-ray cluster, CLJ1001, at $z=2.51$~\citep[][hereafter, W16]{WangT:2016b}, based on CO(1-0) observations with JVLA. CLJ1001 is estimated to have total mass of $M \sim10^{13.9\pm0.2} M_{\odot}$ and virial radius of $R_{\rm 200c} \sim$ 340 kpc based on its X-ray emission and velocity dispersion (W16). Our recent deep narrow-band (NB) imaging further reveals a large number of H$\alpha$-emitters at $z=2.51$ in the cluster, providing further evidence that this is a different structure with respect to the protocluster/large-scale structure found, in the same region of the sky, by~\cite{Casey:2015} at $z = 2.47$. Despite its extended X-ray emission, this cluster is dominated by massive SFGs in the core, which are all detected in CO(1-0), allowing us, for the first time, to probe the gas content and star formation efficiency for a complete sample of massive cluster members (down to $M_{*} > 10^{10.5} M_{\odot}$) at $z > 2$. Throughout the paper, we assume cosmological parameters of $H_{0}$ = 70 km s$^{-1}$ Mpc$^{-1}$, $\Omega_{M}$ = 0.3, and $\Omega_{\Lambda}$ = 0.7. A \cite{Chabrier:2003} initial mass function is adopted to derive stellar masses and SFRs. 
\section{Observations}
\label{Sec:data}

\begin{figure*}[!htb]
\includegraphics[angle=90,scale=0.42]{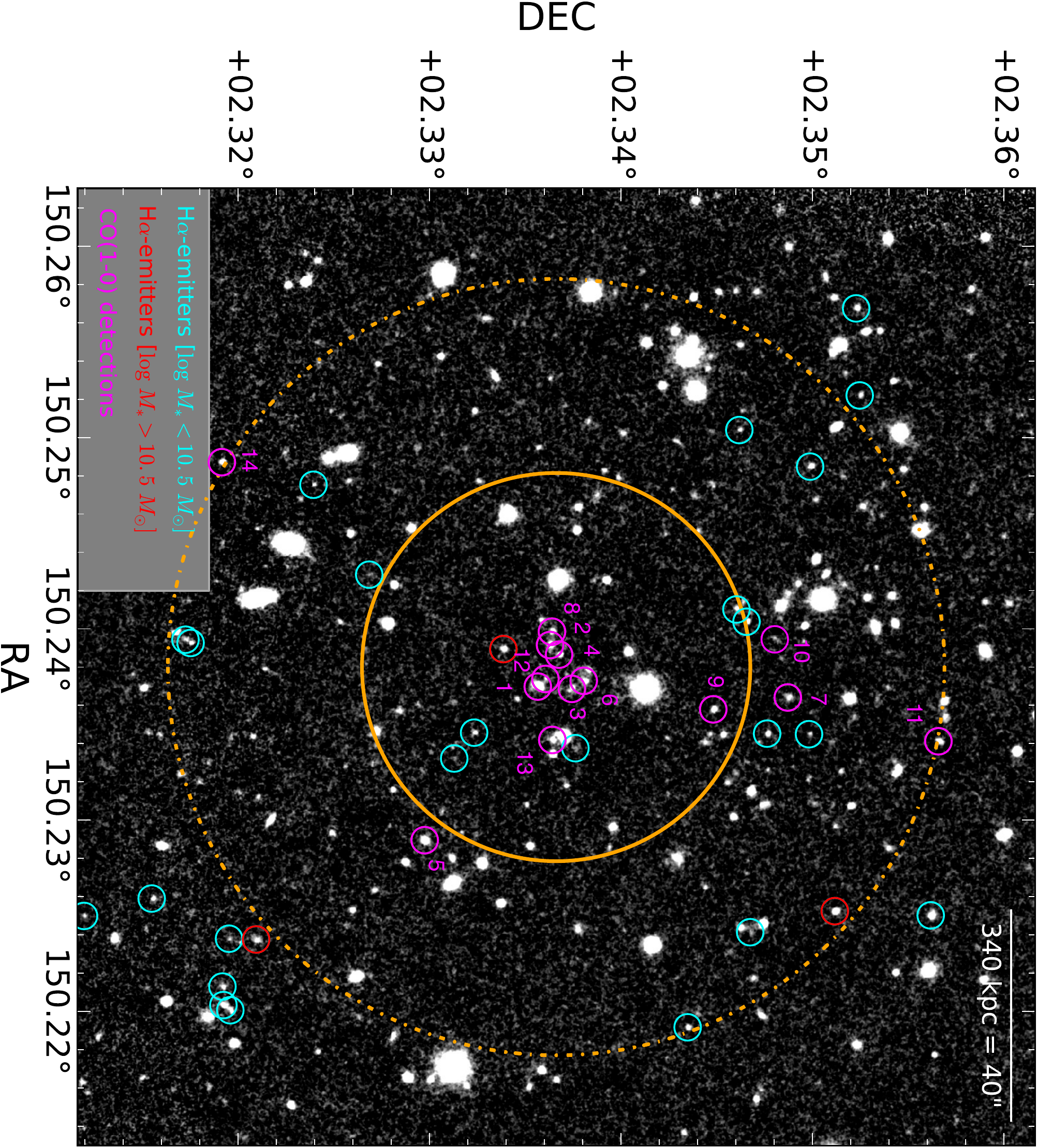}
\includegraphics[scale=0.62]{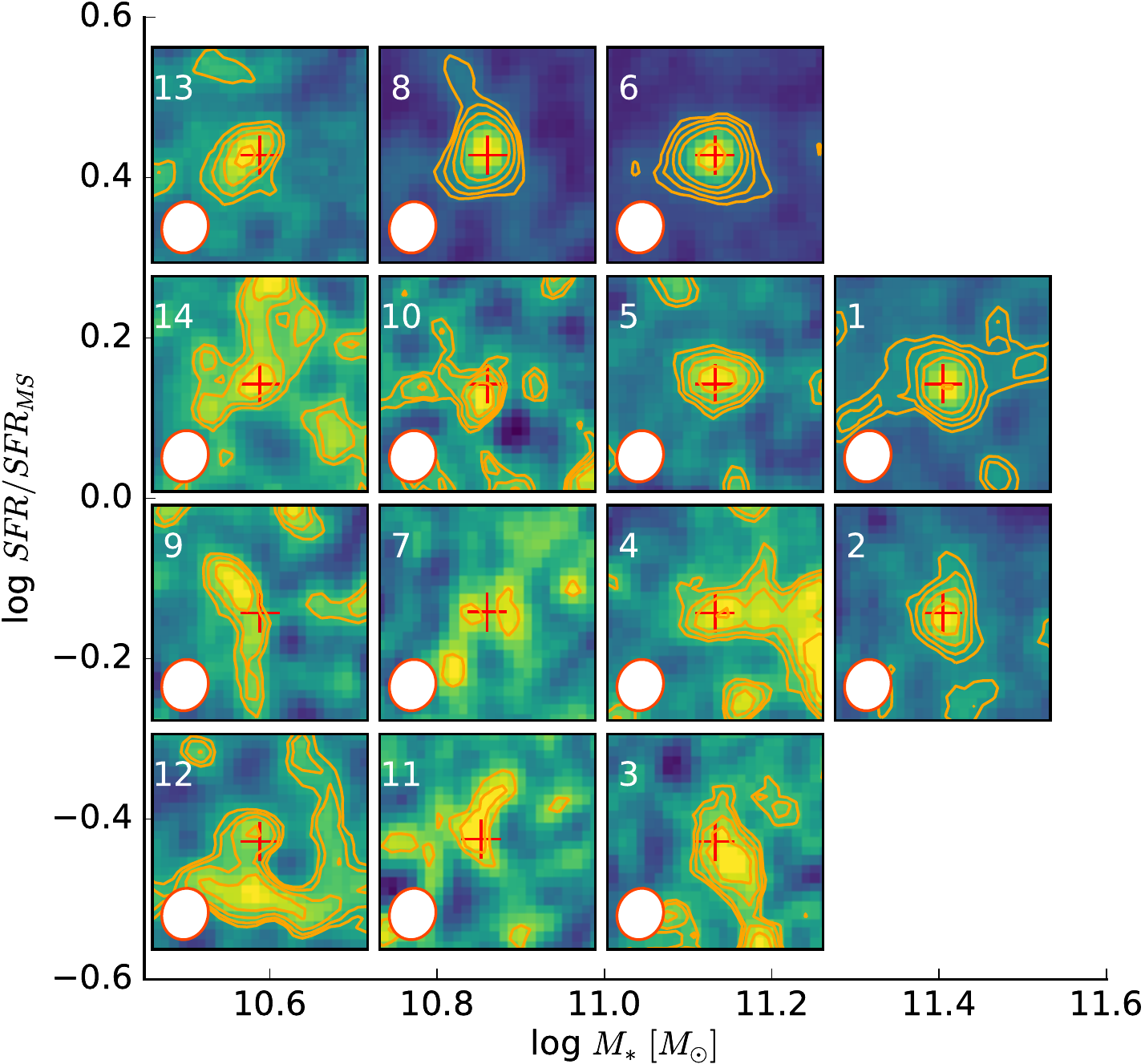}
\caption{
\textbf{Left}: Sky distributions of cluster members around the core of CLJ1001. The background image is the NB image from MOIRCS/Subaru at 2.3 $\mu$m. The cyan and red open circles denote low- and high-mass SF cluster members (H$\alpha$ emitters) at $z = 2.51$  separated at $M_{*} = 10^{10.5} M_{\odot}$. The CO(1-0) detected members all have $M_{*} > 10^{10.5} M_{\odot}$, and are further indicated in magenta. The scale bar indicates the virial radius ($R_{200c}$) of the cluster. The small and large orange circles denote the coverage of the VLA observations, with the diameter corresponding to 1 and 2$\times$ FWHP of the primary beam at 32.878 GHz, respectively. 
\textbf{Right}: The velocity-integrated intensity map (moment-0) of CO $J$=1-0 for the 14 galaxies detected by JVLA. The position of each panel is determined by their stellar mass and SFR (normalized by the SFR of MS galaxies at the same mass). Each panel is 12\arcsec $\times$ 12\arcsec. Contour levels of CO(1-0) starts at $\sqrt2\sigma$  and increase as 2, 2$\sqrt2$, 4, 8, and $16\sigma$. The red cross in each panel indicates the centroid of the stellar emission as determined from the HST/F160W (if available) or NB images. The derived integrated fluxes are presented in Table~\ref{Tab:measurements} .
\label{Fig:sky}}
\end{figure*}

\subsection{JVLA CO(1-0) observations}
Our JVLA observations of CLJ1001 were performed in December 2015 under the program 15B-290 (PI: Tao Wang). 
Part of the data has been already presented in W16 (including example CO(1-0) spectra), which were only used to confirm cluster members. In detail, the observations were carried out in the Ka-band with the D configuration, with an effective frequency coverage of 32.2-33.59 GHz, corresponding to $z\sim 2.43-2.58$ for CO(1-0). The observations were done in excellent weather conditions with Precipitable Water Vapor as low as 2.0 mm and wind speed 0.3-1.5 km/s, therefore the achieved system temperature ($T_{\mathrm{sys}}$) is about or even below 40 K, while typical $T_{\mathrm{sys}}$ is about 50 K at Ka band in winter. The full width half power (FWHP) size of the primary beam is 1.37$\arcmin$ at 32.878 GHz (z=2.506 for CO(1-0)). We observed 3C147 for flux calibration during the full observations,  and a point source J1024-0052 near our target for phase calibration during each scan loop (every $\sim$8 mins). The total integration time is $\sim$13 hours. The data were reduced using the Common Astronomy Software Application (CASA) package~\citep{McMullin:2007} with a standard pipeline. 
We chose 0.5$^{\prime\prime}$ pixels and a spectral resolution of 30 km s$^{-1}$ with a natural weighting scheme for imaging. Image deconvolution was performed with a CLEAN threshold of 3$\sigma$ of each cube. The resulting data cube has synthesized beam size of  $\sim2.88^{\prime\prime} \times 2.52^{\prime\prime}$ with a rms of $\sim 33-40~\mu$Jy beam$^{-1}$ per channel at the phase center. 

\begin{figure*}[!tbh]
\includegraphics[scale=0.75]{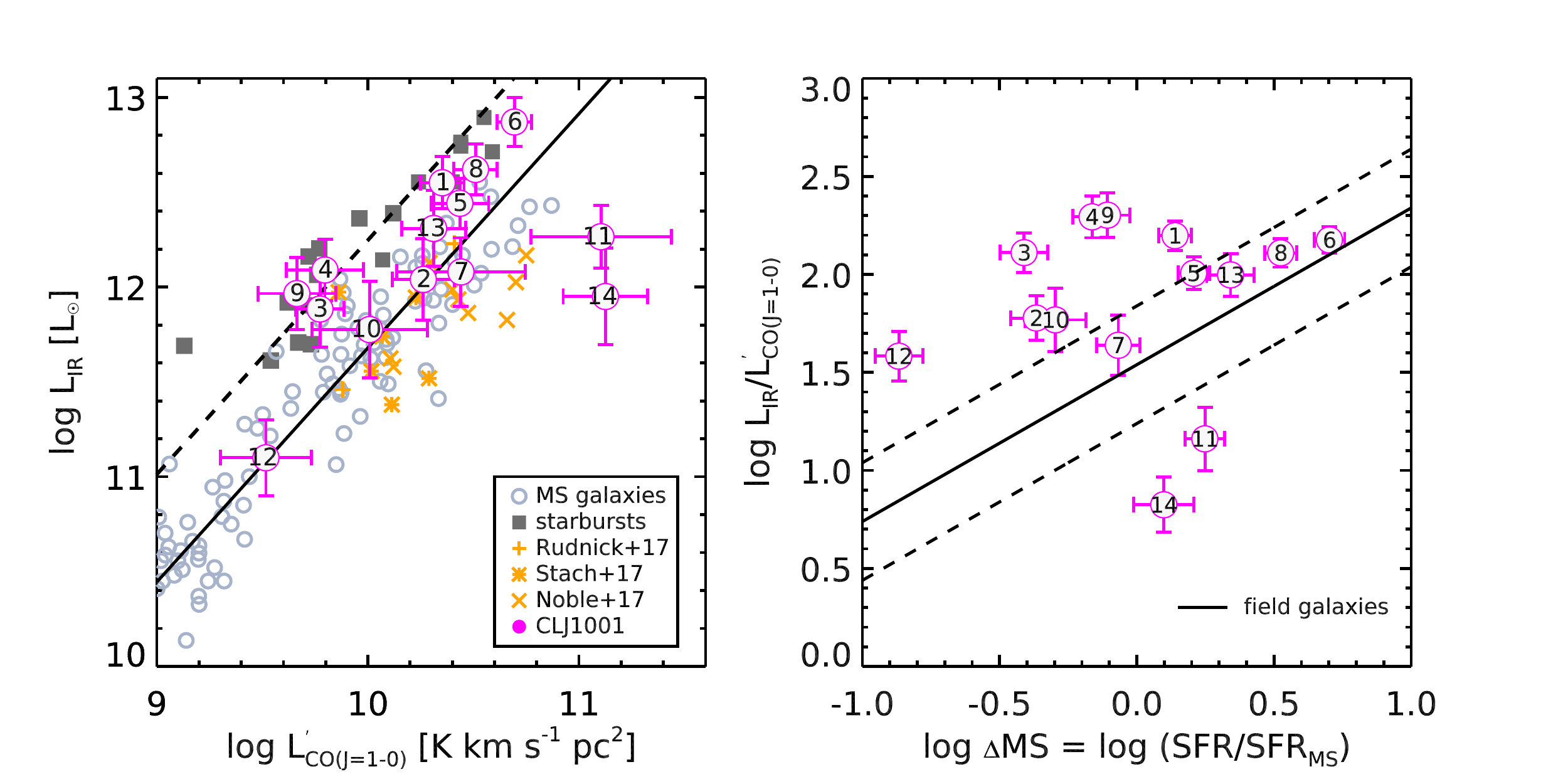}
\caption{\textit{Left}: Infrared luminosities, $L_{\rm IR}$, versus CO(1-0) line luminosities, $L^{\prime}_{\rm CO(1-0)}$, for CO(1-0) detected galaxies in CLJ1001. 
The solid and dashed lines represent the scaling relation for field MS and strong SB galaxies ($\Delta$MS $\sim 10$), respectively~\citep{Sargent:2014}. The field galaxy sample includes both MS and SB galaxies drawn from  the compilation  by \cite{Sargent:2014}, and a sample of $z\sim1$ SB galaxies by \cite{Silverman:2015}. We also include a few recent studies of cluster galaxies at $z \gtrsim 1.5$, which have either CO(2-1)~\citep{Stach:2017,Noble:2017} or CO(1-0)~\citep{Rudnick:2017} measurements. \textit{Right}: Star formation efficiency, as indicated by the ratio between $L_{\rm IR}$ and $L^{\prime}_{\rm CO(1-0)}$, versus $\Delta {\rm MS}$ for galaxies in CLJ1001. The best-fit relation for field galaxies and its associated 1$\sigma$ scatter are also shown~\citep{Magdis:2012b}. A large variety of SFE for cluster galaxies is observed. 
\label{Fig:SFE}}
\end{figure*}

\subsection{Subaru/MOIRCS narrow-band imaging}
To have a complete census of cluster (star-forming) members, we have recently conducted a deep narrow-band (NB) survey towards CLJ1001 with Subaru/MOIRCS. The NB survey employed the ``CO'' filter centered at 2.3$\mu$m to identify H$\alpha$ emitters at $z=2.49-2.52$, combined with the already available deep $K_{s}$-band data in COSMOS from the UltraVista survey~\citep{McCracken:2012,Muzzin:2013a,Laigle:2016}. 
With 4.4 hours of integration, we have detected 49 H$\alpha$ emitters with line flux down to 1.5 $\times 10^{-17}$ erg s$^{-1}$ cm$^{-2}$. This  corresponds to a dust-free SFR of $\sim$ 5 $M_{\odot}$ yr$^{-1}$ at $z$=2.51~\citep{Kennicutt:1998}.
Details of data reduction and star formation properties of these H$\alpha$ emitters will be discussed in a forthcoming paper. Here we only use their positions to search for CO(1-0) line emissions.

\subsection{Extraction of CO(1-0) emitters}

We extract CO(1-0) spectra at the position of the cluster members (H$\alpha$-emitters) out to $2 \times$ the full width half power (FWHP) of the primary beam (PB). 
This approach allows us to detect sources with fainter fluxes and with higher fidelity than a blind search. In total 14 H$\alpha$-emitters are detected with S/N $>$ 3~(Figure~\ref{Fig:sky}). We measured the CO(1-0) line fluxes for each object  by running a 2-D Gaussian fit with CASA (IMFIT) on the velocity-integrated (moment-0) map. The velocity range used to create the moment-0 map of each object was determined so to maximize the signal-to-noise of the detection. During this process the spatial position of the targets was kept fixed based on the coordinates found from the HST/F160W (if available) or NB ancillary images, in order to minimize false detections due to noise fluctuations. 13 out of these 14 galaxies (except ID-14) are also covered by our ALMA band-3 observations, and they are all detected in CO(3-2) at roughly the same velocity~(Wang et al., in preparation). In the case of low S/N with CO(1-0), CO(3-2) data is combined with CO(1-0) to determine the velocity range of the line emission. 
For a sanity check, we have also measured directly their total fluxes in the \textit{uv}-plane with GILDAS\footnote{http://www.iram.fr/IRAMFR/GILDAS}, a procedure that gives consistent results. The measured integrated CO(1-0) line intensities, after primary beam correction, are listed in Table~\ref{Tab:measurements}. The 14 detections include all but one massive galaxy (ID-131651 in W16) with $M_{*} > 10^{10.5} M_{\odot}$ within the PB, which is 
classified as a passive galaxy in W16 based on its rest-frame colors, suggesting that the H$\alpha$ emission mostly likely originates from an (radio)AGN, as further supported by the non-detection of CO(1-0).

\subsection{Molecular gas masses from CO(1-0)}

The use of CO(1-0)  avoids the uncertainty in the CO excitation, and serves as the most extensively used way in obtaining the total molecular gas mass, $M(H_{2})$. The conversion involves the integrated CO emission intensity  (\Lcop) and a conversion factor \aco~\citep{Bolatto:2013}, through $M(H_{2}) = \alpha_{\rm CO} L_{\rm CO}^{\prime}$, with the CO line luminosity \Lcop~derived following \cite{Solomon:2005}. 
We determine \aco~ for the cluster galaxy sample following  \cite{Genzel:2015} and \cite{Tacconi:2018}.
The same mass-metallicity relation used in~\cite{Genzel:2015} is also applied to determine the metallicity for the CO-detected galaxies in our sample, which is close to solar given their large stellar masses. As a result, the derived \aco~is close to the Milky value for this sample (Table~\ref{Tab:measurements}).

\section{Results}
\label{Sec:results}

\begin{figure*}
\centering
\includegraphics[scale=0.78]{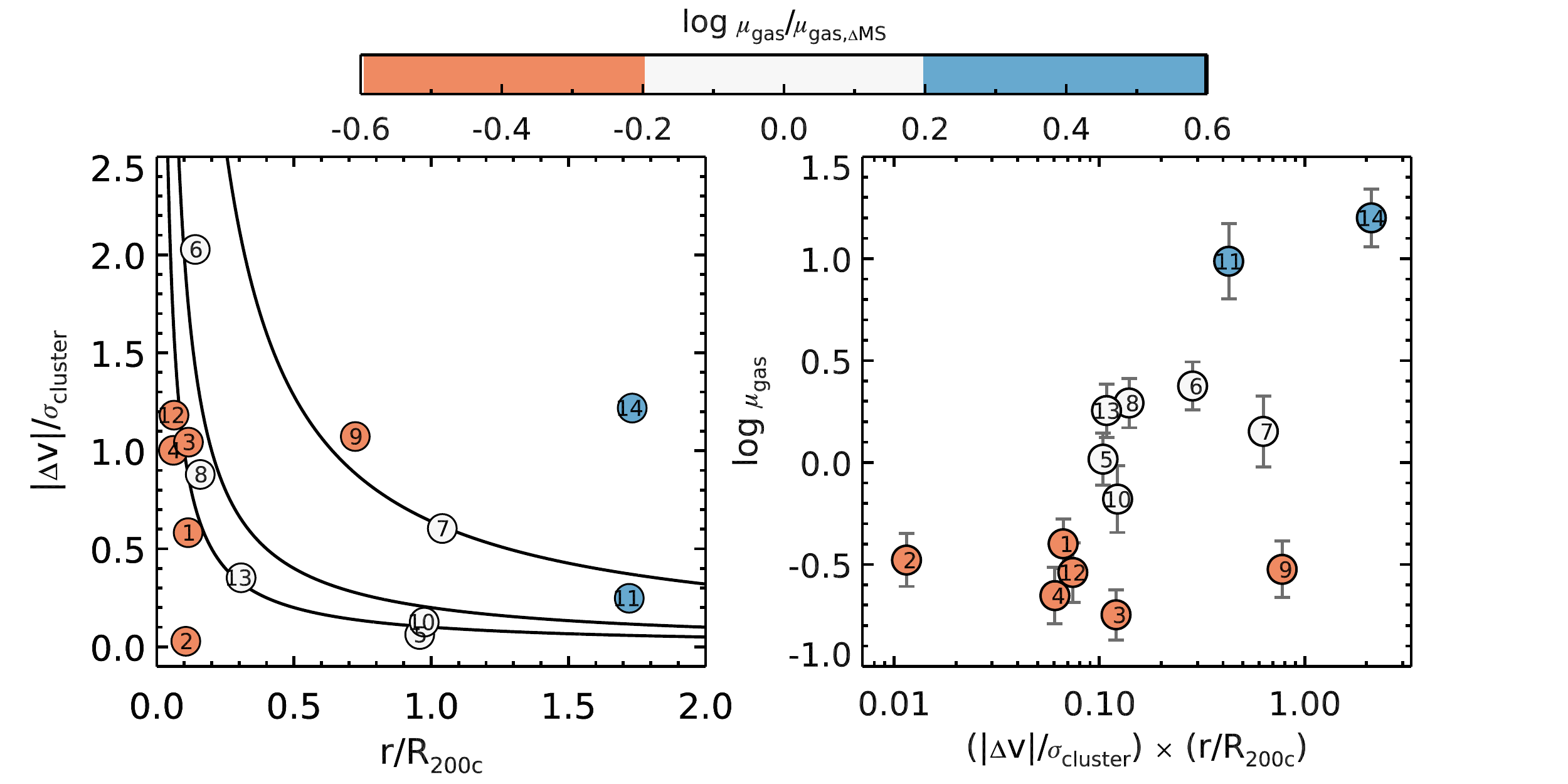}
\caption{Clustercentric-radius dependence of gas fraction as shown in the phase-space diagram (\textit{Left}) and gas fraction versus $k_{b} = (|\Delta v|/\sigma_{\rm cluster}) \times (R/R_{200c})$ plane, which is a proxy for clustercentric radius in 3D (\textit{Right}). Galaxies are color-coded by their gas fraction normalized by the value of field galaxies at the same $\Delta$MS, mass and redshifts~\citep{Tacconi:2018}. Curves of constant $k_{b}$ values with $k_{b} = 0.05, 0.2, 0.64$ are shown in the left panel. A strong clustercentric radius dependence of gas content is revealed, with decreasing gas fraction for galaxies closer to the cluster center, which is true for either absolute or normalized value of gas fraction). 
\label{Fig:phase}}
\end{figure*}

\begin{figure*}
\centering
\includegraphics[scale=0.78]{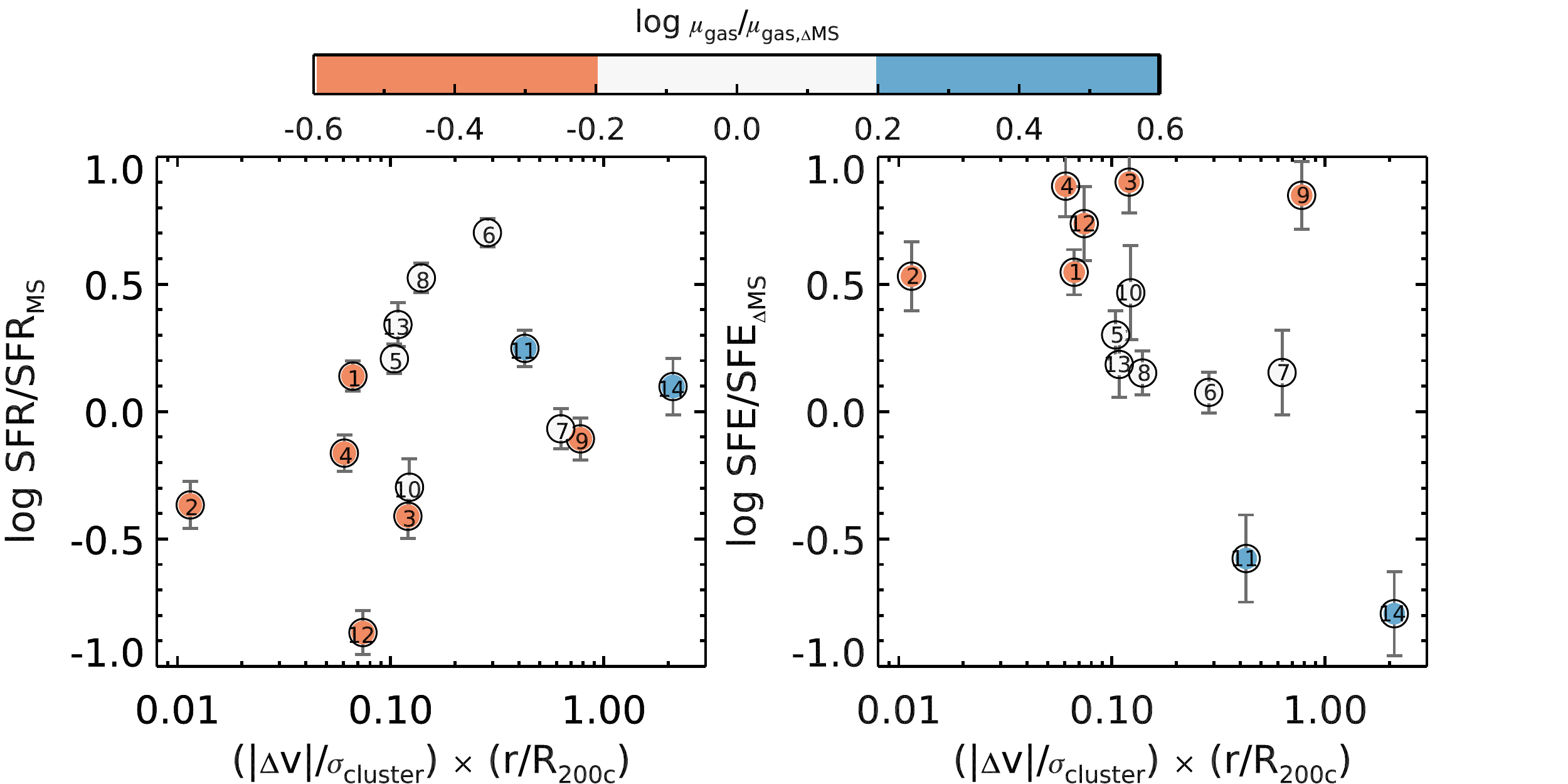}
\caption{Dependence of SFR (\textit{left}) and star formation efficiency (\textit{right}) on the clustercentric radius indicated by $k_{b} = (|\Delta v|/\sigma_{\rm cluster}) \times (R/R_{200c})$. The SFR of each galaxy is normalized by field MS galaxies, while the SFE is normalized by field galaxies at the same $\Delta$MS~\citep{Tacconi:2018, Magdis:2012b}. Galaxies are color-coded by their gas fraction normalized by the field galaxies at the same $\Delta$MS. A general trend of enhanced SFE towards the cluster center is revealed. The low gas fraction yet normal or suppressed SFR of the member galaxies in the cluster center suggest that their enhanced SFE is mainly caused by their deficit of molecular gas (instead of an enhanced SFR).
\label{Fig:fgas_SFE_kb}}
\end{figure*}

\subsection{Star formation efficiency}

The combination of the CO line luminosity (\Lcop), tracing total molecular gas, and the total infrared luminosity (\LIR), tracing newly formed stars, provides a crucial constraint on the star formation efficiency (SFE), with  \LIR/\Lcop $\propto$ SFR/$M_{\rm gas} \equiv$ SFE. Moreover, the use of \LIR/\Lcop~as an approximation of SFE allows us not have to account for the different prescriptions for the CO-to-H$_{2}$ conversion factor, enabling direct comparisons between different samples. As shown in W16~\citep[also see, e.g., ][]{Bussmann:2015}, five cluster members are detected at 870 $\mu$m with ALMA, for which we derive their infrared luminosities, $L_{\rm IR}$, by fitting the full infrared SED. For the other galaxies without 870$\mu$m detection, we derive $L_{IR}$ based on their 24 $\mu$m fluxes~\citep{Muzzin:2013a} and 3 GHz~\citep{Smolcic:2017} radio continuum by using the average infrared SED templates for galaxies at $z\sim 2.5$~\citep{Schreiber:2018a} and FIR-radio relation~\citep{Delhaize:2017}. 
We have verified this approach through comparisons of IR-SED derived and 3 GHz derived \LIR~for four out of the five ALMA-detected sources (excluding ID-4, which is a radio AGN), which are in good agreement. Only one source (ID-12) does not have either 24$\mu$m or 3 GHz detections, for which we derive $L_{\rm IR}$ based on its SFR estimated from extinction-corrected H$\alpha$ following~\cite{Kennicutt:1998}. The best-estimated $L_{\rm IR}$ for cluster members are listed in Table.~\ref{Tab:measurements}.

Figure~\ref{Fig:SFE} presents the comparison of \LIR~and \Lcop~\\
between member galaxies in CLJ1001  and other galaxy populations in high-z clusters ($z \sim 1.5-2$) and field. 
The CLJ1001 galaxies exhibit a large variety in their SFE as traced by \LIR/\Lcop, including members with high, starburst-like SFE and also members with SFE even below MS-like galaxies. To examine whether this large variety in SFE is solely driven by their different star formation modes (starburst versus MS), we further show the variation of  \LIR/\Lcop~as a function of their distance to the MS, which is defined as $\Delta$MS = SFR/SFR$_{\rm MS}$ with SFR$_{\rm MS}$ for each galaxy derived using the MS relation at z=2.5~\citep{Schreiber:2015} at the same stellar mass (the right panel of Figure~\ref{Fig:SFE}), which is nearly identical to the MS relation used in \cite{Tacconi:2018}. Consistent with field galaxies, a general trend towards increasing SFE with enhanced star formation activity (relative to the MS) is observed in CLJ1001. However, in contrast to field galaxies, a large dispersion of SFE  is present for these cluster galaxies. Most prominently, a population of cluster galaxies with MS-like SFR ($|\Delta {\rm MS}| < 0.5$)  exhibit significantly different SFE compared to field galaxies. This suggests that the large variation in SFE for these cluster galaxies is not driven by the variation in SFR, instead, some other mechanisms, most likely related to the dense environment, may play an important role.

\subsection{Clustercentric radius dependence of star formation and molecular gas content}
In order to gain further insights into the origin of the gas and star formation properties of the cluster galaxies, we examine the relation between these properties and their positions in the cluster. Specifically, we employ the line-of-sight velocity versus clustercentric radius phase-space diagram to illustrate the relative distribution of member galaxies within the cluster. The phase-space diagram characterizes the accretion state of cluster member galaxies, which minimizes projection effects of their 2D positions with respect to the cluster center~\citep[see, e.g.,][]{Noble:2013}. 
In this diagram, galaxies that are recently accreted to the cluster tend to have large relative velocities and/or large clustercentric radius, which are offset from the central virialized region. As shown in the left panel of Figure~\ref{Fig:phase}, we observe a clear trend of decreasing gas content ($\mu_{\rm gas} = M_{\rm gas}/M_{*}$) with proximity to the cluster core. This is more clearly illustrated in the right panel of  Figure~\ref{Fig:phase}, in which we plot $\mu_{\rm gas}$ versus $k_{b} = (|\Delta v|/\sigma_{\rm cluster}) \times (R/R_{200c})$ for the cluster galaxies. The parameter $k_{b}$ converts the phase-space diagram into one dimension~\citep{Noble:2013}. Galaxies with lower $k_{b}$ are more closely bounded to the cluster, hence are likely accreted at earlier times. 
Figure~\ref{Fig:phase} reveals clearly a trend that galaxies with high gas fraction (relative to the MS) have entered the cluster more recently than the gas-poor members. This remains true even when the normalized gas fraction, $\mu_{\rm gas}/\mu_{\rm gas, \Delta MS}$, is adopted. While galaxies in the outskirts of the cluster exhibit a large scatter in their gas fraction (compared to field galaxies), galaxies in the cluster center ($k_{b} \lesssim 0.1$) show exclusively a deficit of molecular gas.  
The transition  between the gas-rich and -poor populations takes place at around $k_{b} \sim 0.1$. This rapid transition may suggest that whatever environmental effects are involved, this process must be very efficient in reducing the gas content of cluster galaxies.

\begin{table*}[!tbh]\centering\begin{minipage}{\textwidth} 
\centering
\caption{Physical properties of the CO(1-0)-detected cluster members in CLJ1001}\label{Tab:measurements}
\begin{tabular}{@{}ccccccccccc@{}}
\toprule
\hline
ID  & ID(W16\footnote{IDs are from the $K_{s}$-selected catalog in \cite{Muzzin:2013a}} )  &  $z_{\rm CO}$ & log $M_{*}$   & log $L_{\rm IR}$    & FWHM  & $L^{\prime}_{\rm CO(1-0)}$  &$\alpha_{\rm CO}$ & $M_{\rm gas}$ & $t_{\rm dep}$\\
  & &  &  $M_{\odot}$ & $L_{\odot}$   & 10$^{2}$ km s$^{-1}$ & 10$^{10}$ K km s$^{-1}$ pc$^{2}$ & $M_{\odot}$/(K km s$^{-1}$ pc$^{2}$) & $10^{10} M_{\odot}$ & Gyrs & \\
  \hline
  1  &  130949  & 2.503  & 11.36$\pm$0.15 & 12.55$\pm$0.14   &   5.0$\pm$0.5   & 2.3$\pm$0.2  &  4.06  &   9.2$\pm$0.9  &  0.26\\
  2  &  130901  & 2.507  & 11.35$\pm$0.15 & 12.04$\pm$0.21   &   6.8$\pm$0.9   & 1.8$\pm$0.3  &  4.06  &   7.4$\pm$1.1  &  0.68\\
  3  &  131079  & 2.514  & 11.13$\pm$0.15 & 11.88$\pm$0.20   &   2.8$\pm$1.2   & 0.6$\pm$0.1  &  4.08  &   2.4$\pm$0.3  &  0.32\\
  4  &  130933  & 2.501  & 11.06$\pm$0.15 & 12.09$\pm$0.16   &   6.9$\pm$1.6   & 0.6$\pm$0.1  &  4.08  &   2.6$\pm$0.5  &  0.21\\
  5  &  130359  & 2.508  & 11.03$\pm$0.15 & 12.44$\pm$0.13   &   2.4$\pm$0.4   & 2.7$\pm$0.4  &  4.08  &  11.1$\pm$1.5  &  0.40\\
  6  &  131077  & 2.494  & 10.93$\pm$0.15 & 12.87$\pm$0.13   &   5.5$\pm$0.4   & 4.9$\pm$0.4  &  4.09  &  20.2$\pm$1.7  &  0.27\\
  7  &  132044  & 2.505  & 10.90$\pm$0.15 & 12.08$\pm$0.18   &   6.8$\pm$1.6   & 2.8$\pm$0.8  &  4.09  &  11.3$\pm$3.5  &  0.94\\
  8  &  130891  & 2.513  & 10.83$\pm$0.15 & 12.62$\pm$0.13   &   3.4$\pm$0.2   & 3.2$\pm$0.3  &  4.10  &  13.3$\pm$1.4  &  0.32\\
  9  &  131661  & 2.500  & 10.80$\pm$0.15 & 11.97$\pm$0.19   &   0.9$\pm$0.4   & 0.5$\pm$0.1  &  4.10  &   1.9$\pm$0.3  &  0.20\\
 10  &  131904  & 2.506  & 10.80$\pm$0.15 & 11.78$\pm$0.25   &   6.9$\pm$3.7   & 1.0$\pm$0.3  &  4.10  &  4.2$\pm$1.1  &  0.70\\
 11  &  132627  & 2.506  & 10.73$\pm$0.15 & 12.27$\pm$0.16   &   6.0$\pm$1.9   &12.7$\pm$4.2  &  4.10  &  52.2$\pm$17.4  &  2.83\\
 12  &  130842  & 2.515  & 10.67$\pm$0.15 & 11.10$\pm$0.20   &   0.9$\pm$0.3   & 0.3$\pm$0.1  &  4.11  &   1.4$\pm$0.3  &  1.07\\
 13  & --  & 2.505  & 10.67$\pm$0.15 & 12.31$\pm$0.20   &   5.3$\pm$0.8   & 2.0$\pm$0.3  &  4.11  &   8.4$\pm$ 1.3  &  0.41\\
 14  &  129444  & 2.515  & 10.54$\pm$0.15 & 11.95$\pm$0.25   &   1.4$\pm$0.3   &13.4$\pm$2.7  &  4.12  &  55.0$\pm$11  &  6.17\\
\bottomrule
\end{tabular}
\end{minipage}
\end{table*}

We further present the variation of SFR and SFE as a function of $k_{b}$ in Figure~\ref{Fig:fgas_SFE_kb}, showing that both SFR and SFE also varies with clustercentric radius. While galaxies in the outskirts of the cluster exhibit a large scatter in their SFR, most member galaxies in the center tend to fall below the MS. On the other hand, these galaxies in the center show an significant enhancement in their SFE compared to those in the outskirts and field galaxies. Despite their low gas fraction, their normal or suppressed SFR suggests that their enhanced SFE is mainly caused by their deficit of molecular gas (instead of an enhanced SFR). This indicates that the suppression on the SFR from the dense environment is likely delayed compared to that on the gas content. This high SFE ensures that most of these galaxies will likely consume all their gas in a short time scale.
Their gas depletion time ($t_{\rm dep}$ = 1/SFE) is around $t_{\rm dep} \sim 0.4$Gyrs, a factor of two shorter than field galaxies with the same $\Delta$MS~(Figure~\ref{Fig:SFE}). This time scale is comparable to the cluster dynamical time (approximated by the crossing time), $t_{\rm dyn} \sim R_{200c}/\sigma_{\rm cluster} \sim 0.5$ Gyr, suggesting that most of these cluster galaxies may consume all their gas within a single orbit around the cluster center, and form a passive cluster core by $z \sim 2$.

It should be noted that because the sensitivity of the CO(1-0) observation decreases towards larger radius from the cluster center (phase center), only gas-rich systems can be detected at large radii. However, as shown in Figure~\ref{Fig:sky}, our CO(1-0) detected sample comprise a mass-complete sample of cluster member galaxies, i.e., we are not missing massive star-forming yet gas-poor galaxies up to 2 $\times$ FWHM. Hence our result is not affected by this observational bias.

\section{Discussion and Conclusions}
\label{Sec:summary}

We have obtained CO(1-0) for 14 massive SFGs in the z=2.51 cluster CLJ1001, the largest sample of galaxies within a single cluster with  gas content constraints at $z>2$. These CO-detected galaxies include nearly all the massive SFGs within 2$R_{200c}$, enabling a highly complete census of gas content in massive cluster galaxies. Here we first summarize our main results and then discuss their implications for the formation of massive galaxies in clusters.
Our main results are summarized as follows:

--~Our cluster galaxies exhibit large differences in their star formation activity and gas masses: some are gas-poor, low-SFR, others are gas-rich, starbursting systems. We show that this large variety of properties mainly correlates with the location of galaxies in the cluster (e.g. with their distance from the cluster core and their accretion state). This is particularly clear when considering the phase-space diagram, 
which shows that while galaxies remain relatively gas-rich  when they first enter the cluster, their gas content is rapidly reduced as they approach the cluster center (enter the virial radius).

--~Despite their varieties in gas content and SFR, most cluster galaxies are found to exhibit elevated star formation efficiency with typical gas depletion time of $\sim 400$ Myrs. This gas depletion time is comparable to the dynamical time of the cluster, suggesting that most galaxies may loose their gas (and become quiescent) within a single round-up around the center of the cluster, as further supported by the absence of gas-rich galaxies in the core of this young cluster.

The strong dependence on clustercentric radius of gas content and SFE for these massive SFGs provide evidence that the dense environment plays an important role in shaping the formation/evolution of the most massive cluster galaxies. The significant suppression of molecular gas for all the massive cluster galaxies close to the center (within virial radius) is direct indication that environmental effects helping to stop gas accretion and/or reduce/remove gas content must have taken place. 
Various mechanisms have been proposed in the literature to reduce the gas content of cluster member galaxies such as starvation (namely, further gas accretion is stopped) or ram pressure and tidal stripping (the gas is removed from the galaxies). 
The rapid transition between gas-rich and gas-poor systems in the cluster takes place close to the cluster center with $k_{b} \sim 0.1$, supporting the idea that the main mechanisms involved may be ram pressure and tidal stripping, which happen close to the deep cluster potential~\citep{Treu:2003}. Moreover, the short gas depletion time scale ($\sim$ 0.4 Gyrs, comparable to the dynamical time of the cluster) is also consistent with simulations showing that ram pressure stripping could remove all the gas of cluster members within a single radial orbit around the cluster center~\citep{Cen:2014}. While current observations suffer from relative poor resolutions, future deep, high-resolution observations of both stellar and gas distribution, as well as kinematics would provide more insights into the main environmental mechanisms at work in this young cluster.

\begin{acknowledgements}
We thank the referee for his/her very constructive comments which help improve the quality and clarity of the paper. We thank Zhi-Yu Zhang for useful discussion on data reduction. This paper makes use of the data from Karl G. Jansky Very Large Array, operated by the National Radio Astronomy Observatory. 
This paper makes use of the following ALMA data: ADS/JAO.ALMA\#2016.01155.S and \#2011.0.00539.S. ALMA is a partnership of ESO (representing its member states), NSF (USA) and NINS (Japan), together with NRC (Canada), NSC and ASIAA (Taiwan), and KASI (Republic of Korea), in cooperation with the Republic of Chile. The Joint ALMA Observatory is operated by ESO, AUI/NRAO and NAOJ.
This study was supported by the JSPS Grant-in-Aid for Scientific Research (S) JP17H06130 and the NAOJ ALMA Scientific Research Grant Number 2017-06B. T.W acknowledges the support by the the European Commission through the FP7 SPACE project ASTRODEEP (Ref.No: 312725).

Facilites: VLA, HST, Subaru
\end{acknowledgements}

\end{document}